# Reducing T-Count in quantum string matching algorithm using relative-phase Fredkin gate

Byeongyong Park[1,2], Hansol Noh[1,2], Doyeol Ahn[1,2,*]

[1]*Department of Electrical and Computer Engineering and Center for Quantum Information Processing, University of Seoul, 163 Seoulsiripdae-ro, Dongdaemun-gu, Seoul 02504, Republic of Korea*
[2]*Singularity Quantum Inc, 9506 Villa Isle Drive, Villa Park, CA 92861, USA*
*Corresponding author: dahn@uos.ac.kr

**Abstract**

The string-matching problem, ubiquitous in computer science, can significantly benefit from quantum algorithms due to their potential for greater efficiency compared to classical approaches. The practical implementation of the quantum string matching (QSM) algorithm requires fault-tolerant quantum computation due to the fragility of quantum information. A major obstacle in implementing fault-tolerant quantum computation is the high cost associated with executing T gates. This paper introduces the relative-phase Fredkin gate as a strategy to notably reduce the number of T gates (T-count) necessary for the QSM algorithm. This reduces the T-count from $14N^{3/2}log_2 N - O(N^{3/2})$ to $8N^{3/2}log_2 N - O(N^{3/2})$, where $N$ represents the size of the database to be searched. Additionally, we demonstrate that our method is advantageous in terms of other circuit costs, such as the depth of T gates and the number of CNOT gates. This advancement contributes to the ongoing development of the QSM algorithm, paving the way for more efficient solutions in the field of computer science.

**Keywords:** quantum circuit, quantum string matching, T-count, fault-tolerant quantum computing, relative phase Fredkin gate, Grover's search algorithm

# 1 Introduction

Grover's quantum search algorithm is renowned for providing a quadratic speedup in solving unstructured search problems [1, 2]. A notable application of this speedup is in addressing the string-matching problem [3, 4], where the goal is to find a pattern $P$ of length $M$ within a given string $L$ of length $N$. This problem is fundamental in computer science due to its wide-ranging applications [5], including information retrieval [6], plagiarism detection [7], text mining [8], intrusion detection [9, 10], language translation [11], and DNA sequence analysis [12−15], among others.

A significant challenge in implementing the quantum string matching (QSM) algorithm is the requirement for fault-tolerant quantum computation due to the inherent fragility of quantum information. In fault-tolerant quantum computations, circuits are constructed using a universal and fault-tolerant gate set. The Clifford + T gate set is commonly employed in various leading fault-tolerant schemes [16, 17]. Within these approaches, Clifford gates are relatively straightforward to implement, often through transversal methods. In contrast, T gates are resource-intensive and require techniques like magic state distillation, which significantly increase the implementation costs [18−21]. Consequently, optimizing the number of T gates (T-count) is crucial for the efficient execution of large-scale quantum algorithms, including QSM.

Several studies have developed QSM algorithms [3, 4, 22], with the most notable contribution being the work of Niroula and Nam [4]. Their work is particularly significant because it provides an explicit quantum circuit implementation of the QSM algorithm. In their circuit design, they introduce a component called the 'cyclic operator,' which plays an essential role in constructing the Grover's operator specific to the QSM. The construction of these operators in the QSM circuit is the primary factor that increases the T-count within the QSM algorithm.

In this paper, we propose a method to reduce the T-count for the QSM circuit, based on the circuit design proposed by Niroula and Nam [4]. The QSM circuit in Niroula and Nam's work requires approximately $14N^{3/2} \log_2 N$ T gates, the majority of which arise from the synthesis of controlled-SWAP (Fredkin) gates in cyclic operators. They synthesized each Fredkin gate into Clifford + T gates using seven T gates (See Fig. 1(a)). As part of our strategy to reduce the T-count, we introduce a new concept: the relative-phase Fredkin gate, inspired by the relative-phase Toffoli gate [23]. We synthesize a relative-phase Fredkin gate that contains only four T gates (See Fig. 1(b)) and prove that each Fredkin gate in the QSM circuit can be replaced with any relative-phase Fredkin gate. As a result, we reduce the leading-order term of the T-count for QSM to $8N^{3/2} \log_2 N$. Additionally, we demonstrate that our approach provides advantages in terms of reducing other circuit costs, such as the depth of T gates (T-depth) and the number of CNOT gates (CNOT-count).

The remainder of this paper is structured as follows. Section 2 provides background information, focusing on Grover's algorithm and the QSM circuit described in Ref. [4]. Section 3 details the construction of the QSM circuit using relative-phase Fredkin gates and demonstrates that the QSM circuit functions correctly. Section 4 presents the cost of the QSM algorithm, focusing on T-count. Section 5 concludes by discussing the results and their implications.

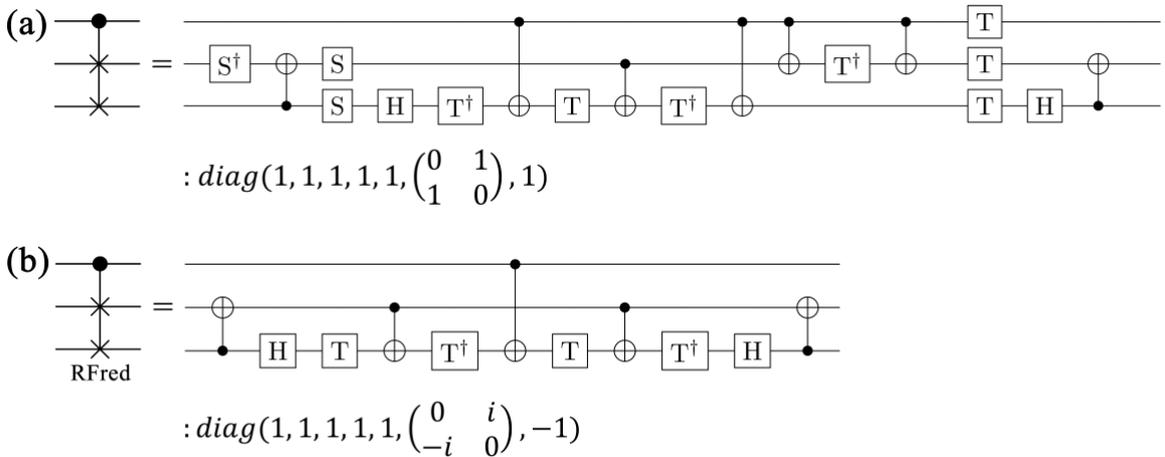

**Fig. 1** Decomposition of the Fredkin gate and the relative-phase Fredkin gate into the Clifford + T gate set. (a) Decomposition of the Fredkin gate used in Ref. [4], which requires seven T gates. (b) Decomposition of the relative-phase Fredkin gate used in this study, which requires four T gates. Implementing the QSM algorithm with the relative-phase Fredkin gate significantly reduces the T-count, a critical factor in minimizing space-time costs in fault-tolerant quantum computing.

## 2 Backgrounds

### 2.1 Grover's algorithm

Grover's algorithm is a quantum algorithm that offers a quadratic speedup over classical search algorithms in unstructured search problems [1, 2]. This section briefly explains Grover's algorithm using general quantum amplitude amplification [2].

Consider a Boolean function $f: \{0, 1, \ldots, 2^n - 1\} \to \{0, 1\}$. Within an $n$-qubit system, two subspaces, $G$ and $B$, are defined as follows: $G \equiv span(\{|x\rangle \mid f(x) = 1\})$ and $B \equiv span(\{|x\rangle \mid f(x) = 0\})$. Note that $I_n = P_G + P_B$, where $P_G$ and $P_B$ are projectors onto $G$ and $B$, respectively. Given a unitary operator $A$ acting on the $n$-qubit system, $A|0\rangle_n \equiv |\psi\rangle$ can be expressed as a linear combination of $P_G|\psi\rangle$ and $P_B|\psi\rangle$ as shown in Eq. (1). Here, the states $|\varphi_g\rangle$ and $|\varphi_b\rangle$ are normalized states of $P_G|\psi\rangle$ and $P_b|\psi\rangle$, respectively.

$$A|0\rangle_n \equiv |\psi\rangle = P_G|\psi\rangle + P_B|\psi\rangle = \sin\theta |\varphi_g\rangle + \cos\theta |\varphi_b\rangle \tag{1}$$

In this study, we call the operator $A$ the initialization operator.

The primary goal of Grover's algorithm is to find (or measure) the state in $G$ with high probability. The operators utilized in Grover's algorithm are defined in Eqs. (2) to (5):

$$Q \equiv -R_\psi R_g \tag{2}$$

$$R_\psi \equiv I - 2|\psi\rangle\langle\psi| \equiv A R_0 A^{-1} \tag{3}$$

$$R_0 \equiv I - |0\rangle_n\langle 0| \tag{4}$$

$$R_g \equiv I - 2|\varphi_g\rangle\langle\varphi_g| \tag{5}$$

We refer to operator $Q$ as Grover's operator. The process of Grover's algorithm can be delineated as follows: First, an $n$-qubit is set to state $|0\rangle_n$, after which the operator $A$ is applied to the state $|0\rangle_n$. Next, $Q^k$ is applied to the state $A|0\rangle_n = |\psi\rangle$, where $k = \lfloor \pi/4\theta \rfloor$. Finally, a measurement is conducted on the $n$-qubit state. The post measurement state becomes a state in $G$ with a probability higher than $max(1 - \sin^2\theta, \sin^2\theta)$ [2].

## 2.2 Quantum string matching algorithm

This section describes the QSM algorithm and its circuit representation described in Ref. [4]. String matching aims to locate a pattern $P$ of length $M$ in a given string $L$ of length $N$. For simplicity of analysis, we assume $P$ and $L$ to be binary digit strings and $N = 2^n$ throughout this study.

The core subcircuit of the algorithm is the cyclic operator $C$. The cyclic operator $C$ on the $(n + N)$-qubit system is defined as Eq. (6), where $|k\rangle$ is a computational basis state, the additions are modulo $N$ additions, and each $x_i$ is a binary number 0 or 1.

$$C|k\rangle_n|x_0 x_1 x_2 \ldots x_{N-1}\rangle_N = |k\rangle_n|x_{0+k} x_{1+k} x_{2+k} \ldots x_{N-1+k}\rangle_N \tag{6}$$

To construct the circuit of this operator, the cyclic operator $C$ is divided into operators $C_k$, which are defined as follows.

$$C_k|l\rangle_n|x_0 x_1 x_2 \ldots x_{N-1}\rangle_N = \begin{cases} |l\rangle_n|x_0 x_1 x_2 \ldots x_{N-1}\rangle_N, & l \neq k, \\ |l\rangle_n|x_{0+k} x_{1+k} x_{2+k} \ldots x_{N+k-1}\rangle_N, & l = k, \end{cases} \tag{7}$$

where $|l\rangle$ is a computational basis state. Then, the cyclic operator $C$ can then be rewritten as in Eq. (8).

$$C = C_{2^0} C_{2^1} C_{2^2} \ldots C_{2^{n-1}} \tag{8}$$

Constructing a $C_{2^j}$ operator requires maximum of $N - 1$ controlled-SWAP (Fredkin) gates. The process of constructing the $C_{2^j}$ operator is divided into $n$ stages. In the first stage, $N/2$ digits are moved to the right place using $N/2$ Fredkin gates. In the second stage, $N/4$ digits are moved to the right place using $N/4$ Fredkin gates, and so on. The total number of Fredkin gates used to construct the $C_{2^j}$ operator is at most $\frac{N}{2} + \frac{N}{4} \ldots + 1 = N - 1$ (See Fig. 2). Optionally, the Fredkin gates at each stage of constructing a $C_{2^j}$ operator can be parallelized using $\frac{N}{2} - 1$ ancilla qubits and fan-out operations with $N - 2$ CNOT gates. The exact number of Fredkin gates required to construct the cyclic operator $C$ can be found in Appendix A.

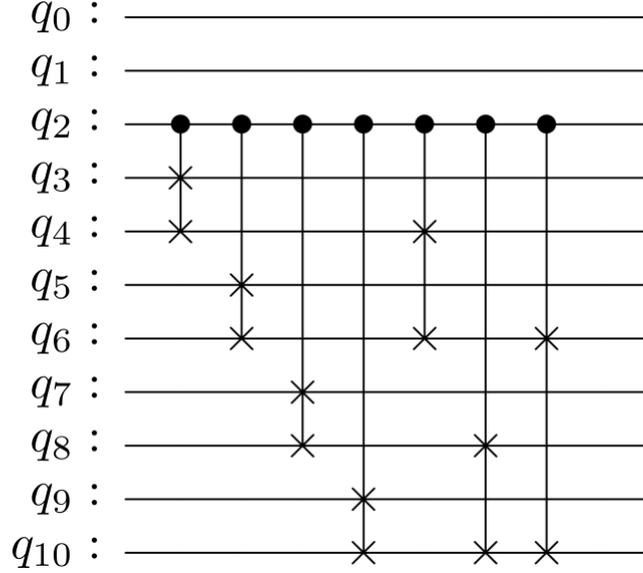

**Fig. 2** Construction of a $C_{2^0}$ operator. In this figure, we show a $C_{2^0}$ circuit when $N = 8$.

To implement Grover's algorithm, the initialization operator $A$ such that $A|0\rangle_{n+N+M} = |\psi\rangle$ need to be defined. The process of applying $A$ to state $|0\rangle_{n+N+M}$ is as follows (See Fig. 3(a)):

(1) Prepare three quantum registers with $n$, $N$, and $M$ qubits. We call the $n$-qubit register the first register, $N$-qubit register the second register, and $M$-qubit register the third register. All the states were initialized to state $|0\rangle$.

(2) Apply $H^{\otimes n}$ to the first register and apply $X$ gates to encode $L$ and $P$ in the second and third registers, respectively. Then, the entire state becomes state $|\psi_1\rangle$ in Eq. (9). $|L\rangle_N$ and $|P\rangle_M$ in Eq. (9) are defined in Eqs. (10) and (11), respectively, where each $l_j$ and $p_j$ is a binary number.

$$|\psi_1\rangle = \frac{1}{\sqrt{N}} \sum_{k=0}^{N-1} |k\rangle_n |L\rangle_N |P\rangle_M \qquad (9)$$

$$|L\rangle_N = |l_0 l_1 l_2 \dots l_{N-1}\rangle \qquad (10)$$

$$|P\rangle_M = |p_0 p_1 p_2 \dots p_{M-1}\rangle \qquad (11)$$

(3) Apply cyclic operator $C$ to the first and second registers. Subsequently, the state evolves into state $|\psi_2\rangle$ in Eq. (12), where the additions are modulo $N$ additions.

$$|\psi_2\rangle = \frac{1}{\sqrt{N}} \sum_{k=0}^{N-1} |k\rangle_n |l_{0+k} l_{1+k} l_{2+k} \dots l_{N-1+k}\rangle |p_0 p_1 p_2 \dots p_{M-1}\rangle \quad (12)$$

(4) Apply $M$ CNOT gates to first $M$ qubits in the second register and to all qubits in the third register. For the CNOT operations, each qubit in the second register serves as the control qubit, and each qubit in the third register serves as the target qubit (See Fig. 3(b)) Subsequently, the state evolves into state $|\psi\rangle$ as described in Eq. (13).

$$|\psi\rangle = \frac{1}{\sqrt{N}} \sum_{k=0}^{N-1} |k\rangle_n |l_{0+k} l_{1+k} \dots l_{N+k-1}\rangle |(p_0 \oplus l_{0+k})(p_1 \oplus l_{1+k}) \dots (p_{M-1+k} \oplus l_{M-1+k})\rangle \quad (13)$$

In the remainder of this study, we use the notation in Eqs. (14) and (15).

$$|l_{0+k} l_{1+k} \dots l_{N+k-1}\rangle \equiv |L_k\rangle_N \quad (14)$$

$$|(p_0 \oplus l_{0+k})(p_1 \oplus l_{1+k}) \dots (p_{M-1+k} \oplus l_{M-1+k})\rangle \equiv |P_k\rangle_M \quad (15)$$

We define the state in $G$ as the state in which the third register is in state $|0\rangle_M$. If Grover's algorithm is implemented using operator $A$ of Fig. 3(a), we can find pattern $P$ in string $L$.

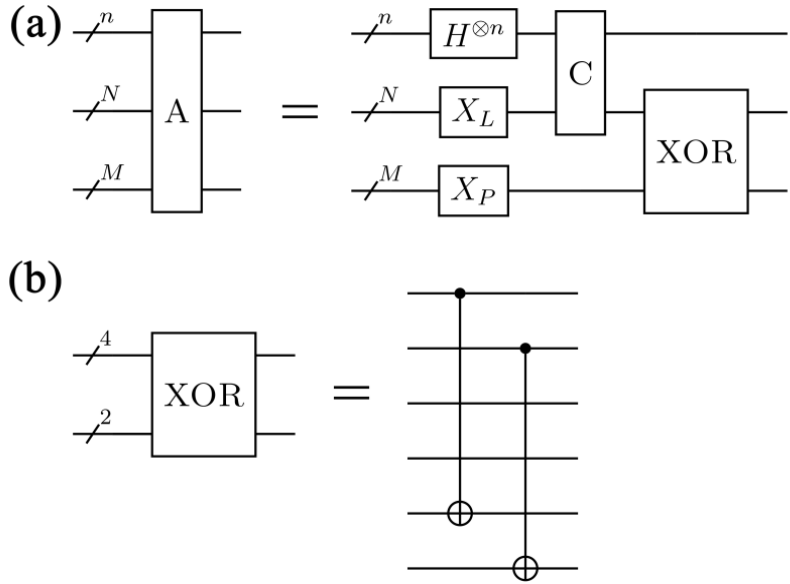

**Fig. 3** Construction of the initialization operator $A$. (a) Structure of the initialization operator $A$. $X_L$ and $X_P$ represent $X$ gates to encode $L$ and $P$, respectively. $C$ represents the cyclic operator. $XOR$ transforms state $|\psi_2\rangle$ in Eq. (12) into state $|\psi\rangle$ in Eq. (13). (b) Example of constructing an $XOR$ operation when $N = 4$ and $M = 2$.

As described in the previous section, Grover's operator $Q$ is $AR_0A^{-1}R_g$. $R_0$ and $R_g$ can be constructed using $X$ gates and a multi-qubit controlled-$Z$ gate. A $k$-qubit controlled-$Z$ gate is decomposed using two $H$ gates and a $k$-qubit Toffoli gate. In Ref. [4], multi-qubit Toffoli gates are decomposed using relative-phase Toffoli gates in Ref. [23]. The calculations regarding the size and cost of $R_0$, which is not explicitly presented in Ref. [4], can be found in Appendix B.

## 3 Proposed QSM circuit: design and validation

First, we define the relative-phase Fredkin gate analogously to the relative-phase Toffoli gate [23]. The Fredkin gate, also known as the controlled-SWAP gate, has a matrix representation on a computational basis given by $diag(1,1,1,1,1,\begin{pmatrix} 0 & 1 \\ 1 & 0 \end{pmatrix},1)$. We extend this to define the relative-phase Fredkin gate, which has a matrix representation of $diag(z_0, z_1, z_2, z_3, z_4, \begin{pmatrix} 0 & z_6 \\ z_5 & 0 \end{pmatrix}, z_7)$, where each $z_i$ is an arbitrary complex number with norm 1.

A Fredkin gate can be constructed using a Toffoli gate and two CNOT gates (see Fig. 4(a)) [24]. Similarly, a relative-phase Fredkin gate can be constructed using a relative-phase Toffoli gate and two CNOT gates (see Fig. 4(b)). This construction can be easily demonstrated through a straightforward matrix multiplication:

$$diag\left(\begin{pmatrix} 1 & 0 & 0 & 0 \\ 0 & 0 & 0 & 1 \\ 0 & 0 & 1 & 0 \\ 0 & 1 & 0 & 0 \end{pmatrix}, \begin{pmatrix} 1 & 0 & 0 & 0 \\ 0 & 0 & 0 & 1 \\ 0 & 0 & 1 & 0 \\ 0 & 1 & 0 & 0 \end{pmatrix}\right)$$

$$\cdot diag\left(z_0, z_1, z_2, z_3, z_4, z_5, \begin{pmatrix} 0 & z_7 \\ z_6 & 0 \end{pmatrix}\right)$$

$$\cdot diag\left(\begin{pmatrix} 1 & 0 & 0 & 0 \\ 0 & 0 & 0 & 1 \\ 0 & 0 & 1 & 0 \\ 0 & 1 & 0 & 0 \end{pmatrix}, \begin{pmatrix} 1 & 0 & 0 & 0 \\ 0 & 0 & 0 & 1 \\ 0 & 0 & 1 & 0 \\ 0 & 1 & 0 & 0 \end{pmatrix}\right)$$

$$= diag(z_0, z_1, z_2, z_3, z_4, \begin{pmatrix} 0 & z_6 \\ z_5 & 0 \end{pmatrix}, z_7) \blacksquare \qquad (16)$$

Using this method, as shown in Fig. 1(b), we constructed a relative-phase Fredkin gate from the relative-phase Toffoli gate described in Ref. [23]. The relative-phase Fredkin gate in Fig. 1(b) has a matrix representation of $diag(1,1,1,1,1,\begin{pmatrix}0 & i\\ -i & 0\end{pmatrix},-1)$.

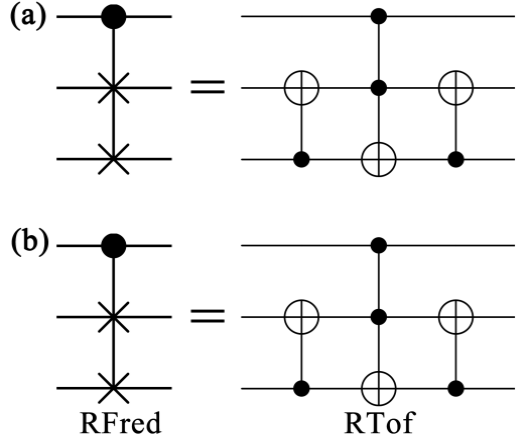

**Fig. 4** Decomposition of Fredkin gate and relative-phase Fredkin gate. (a) Construction of a Fredkin gate using a Toffoli gate and two CNOT gates reported in Ref. [24]. (b) Construction of a relative-phase Fredkin gate using a relative-phase Toffoli gate and two CNOT gates.

Next, we prove that any relative-phase Fredkin gate can replace the Fredkin gates used to construct the cyclic operator $C$ in the QSM algorithm.

*Lemma 1.* Any quantum circuit $X_r$ composed solely of relative-phase Fredkin gates can be represented by Eq. (17), where $X$ is a quantum circuit in which all relative-phase Fredkin gates in $X_r$ are replaced with Fredkin gates, and $D_r$ and $D_l$ are circuits with diagonal matrix representation.

$$X_r = X \cdot D_r = D_l \cdot X \tag{17}$$

*Proof.* It is sufficient to verify Eq. (18), where $RFred$ represents an arbitrary relative-phase Fredkin gate, $Fred$ represents a Fredkin gate, and $D_1$ and $D_2$ represent circuits with a diagonal matrix representation.

$$RFred = Fred \cdot D_1 = D_2 \cdot Fred \tag{18}$$

Basic matrix multiplication can prove Eq. (18):

$$diag(z_0, z_1, z_2, z_3, z_4, \begin{pmatrix} 0 & z_6 \\ z_5 & 0 \end{pmatrix}, z_7)$$

$$= diag\left(1, 1, 1, 1, 1, \begin{pmatrix} 0 & 1 \\ 1 & 0 \end{pmatrix}, 1\right) \cdot diag(z_0, z_1, z_2, z_3, z_4, z_5, z_6, z_7)$$

$$= diag(z_0, z_1, z_2, z_3, z_4, z_6, z_5, z_7) \cdot diag\left(1, 1, 1, 1, 1, \begin{pmatrix} 0 & 1 \\ 1 & 0 \end{pmatrix}, 1\right) \blacksquare \quad (19)$$

*Theorem 1.* Even if arbitrary relative-phase Fredkin gates replace the Fredkin gates in constructing the cyclic operator $C$ in the QSM algorithm described in Ref. [4], the algorithm yields the same results.

*Proof.* Assume that we replace all Fredkin gates in cyclic operator $C$ with relative-phase Fredkin gates. We express operators $A$ and $C$ with relative-phase Fredkin gates as $A'$ and $C'$. By Lemma 1, $C'$ can be written as $C \cdot D$, where $D$ is an $(n + N)$-qubit operator with a diagonal matrix representation. Subsequently, $A|0\rangle_n|0\rangle_N|0\rangle_M$ and $A'|0\rangle_n|0\rangle_N|0\rangle_M$ evolve into the states in Eqs. (20) and (21), where each $z_k$ is a complex number with norm 1.

$$A|0\rangle_n|0\rangle_N|0\rangle_M = |\psi\rangle = \frac{1}{\sqrt{N}} \sum_{k=0}^{N-1} |k\rangle_n|L_k\rangle_N|P_k\rangle_M \equiv sin\theta|\varphi_g\rangle + cos\theta|\varphi_b\rangle \quad (20)$$

$$A'|0\rangle_n|0\rangle_N|0\rangle_M = |\psi'\rangle = \frac{1}{\sqrt{N}} \sum_{k=0}^{N-1} z_k|k\rangle_n|L_k\rangle_N|P_k\rangle_M \equiv sin\theta'|\varphi'_g\rangle + cos\theta|\varphi'_b\rangle \quad (21)$$

Note that $\theta = \theta'$ because states $|k\rangle_n|L_k\rangle_N|P_k\rangle_M$ with different $k$ values are orthonormal. Therefore, Grover's algorithm using $A'$ gives the same results as the case using $A$ $\blacksquare$

Based on Theorem 1, we can replace every Fredkin gate in the QSM algorithm with the relative-phase Fredkin gate of Fig. 1(b). Seven T gates are required to synthesize a Fredkin gate into Clifford + T gates (See Fig. 1(a)). However, only four T gates are required to synthesize the relative-phase Fredkin gate of Fig. 1(b). Therefore, with this replacement, the cyclic operator $C'$ is decomposed into Clifford + $T$ gates using $4(Nlog_2N - N + 1)$ T gates. This results in reducing the leading-order term of the T-count in the QSM algorithm from $14N^{3/2} \log_2 N$ to $8N^{3/2} \log_2 N$, assuming that $N^{1/2}$ Grover iterations (the number of executions of the Grover operator) are implemented.

To validate the proposed QSM circuit, we simulated the QSM algorithm using the quantum simulator in Qiskit [25]. As shown in Fig. 5, the proof-of-principle results exhibit excellent agreement with the expected outcomes of Grover's search algorithm. Our designed circuit targeted the pattern '11' within the data sequence '00110000'. We varied the number of Grover iterations from 0 to 9 and compared the probabilities of finding the correct answer between the theoretically ideal Grover's search algorithm and the QSM algorithm implemented with the relative-phase Fredkin gate. Each QSM simulation was executed 10,000 times to estimate the probabilities.

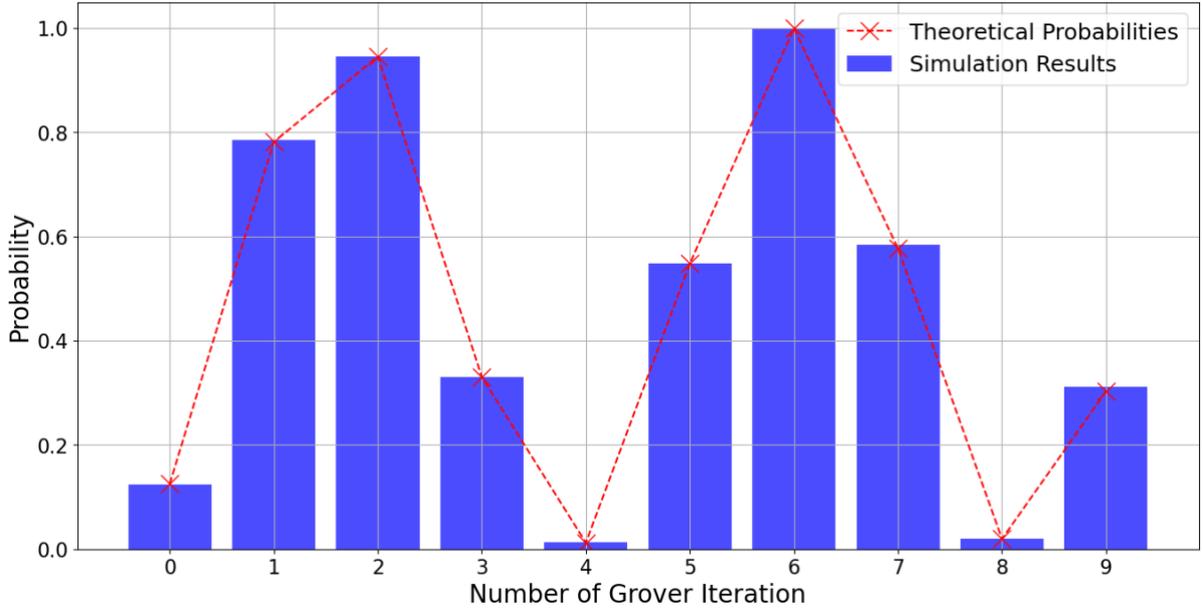

**Fig. 5** Comparison of the theoretical probabilities of Grover's search and the results of the QSM simulation with relative-phase Fredkin gates. Each QSM simulation was executed 10,000 times to estimate the probabilities. The designed circuit was intended to identify the pattern '11' within the data sequence '00110000'. The x-axis shows the number of Grover iterations, and the y-axis displays the corresponding probabilities. The red line plot with "x" markers represents the theoretical probabilities of Grover's search algorithm functioning correctly, while the blue bar graph indicates the simulation results of the QSM algorithm with relative-phase Fredkin gates. The simulation data aligns closely with the theoretical expectations, demonstrating the correctness of the QSM algorithm with relative-phase Fredkin gates.

## 4 Results

In this section, we present the T-count required for the implementation of the QSM algorithm with relative-phase Fredkin gates. When Grover's operator is repeatedly implemented $N^{1/2}$ times, the leading order term of the total T-count for the QSM algorithm is reduced from $14N^{3/2}\log_2 N$ to $8N^{3/2}\log_2 N$. Additionally, we observe further T-count reductions in each iteration of $A'R_0 A'^{-1}$, where the primed operators refer to those using the relative-phase Fredkin gates. To construct $C'$, we sequentially combine $C'_1, C'_2, C'_4$ ... in sequence. For $C'^{-1}$, we combine the inverse gates of the gates constituting $C'$ in reverse order. Then, as shown in Fig. 6, there is a T-count reduction around $R_0$. Each time the structure of the $A'R_0A'^{-1}$ appears in the QSM circuit, the T-count is reduced by $2N$ compared to the case of the QSM algorithm with Fredkin gates.

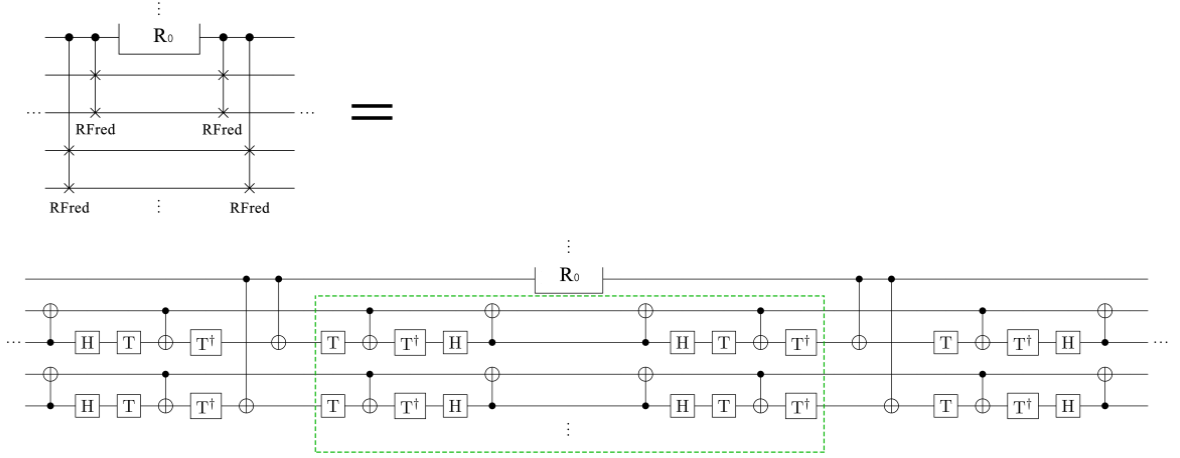

**Fig. 6** Circuit of $A'R_0A'^{-1}$. The gates within the dashed green box are cancelled out, resulting in a reduction of $2N$ T-count.

The introduction of the proposed relative-phase Fredkin gate for the QSM algorithm also benefits the T-depth by allowing the parallelization of the T gates without additional ancilla qubits for the construction of the cyclic operator. The QSM algorithm described in Ref. [4] parallelizes Fredkin gates in the cyclic operator by using $\frac{N}{2} - 1$ ancilla qubits and fan-out operations with CNOT gates, which leads to the parallelization of the T gates. In contrast, the use of relative-phase Fredkin gates allows for automatic parallelization of T gates without the need for fan-out operations or additional ancilla qubits, as depicted in Fig. 7. Compared to the Fredkin gate in Fig. 1(a) with a T-depth of five, the relative-phase Fredkin gate in Fig. 1(b) has a T-depth of four. Consequently, the T-depth required for the cyclic operator is reduced to 4/5 of the result in Ref. [4] without additional ancilla qubits. The use of relative-phase Fredkin gates also offers advantages in terms of other circuit costs, such as the number of CNOT gates. In Table I, the circuit cost of the QSM algorithms proposed in this paper is compared against those previously reported.

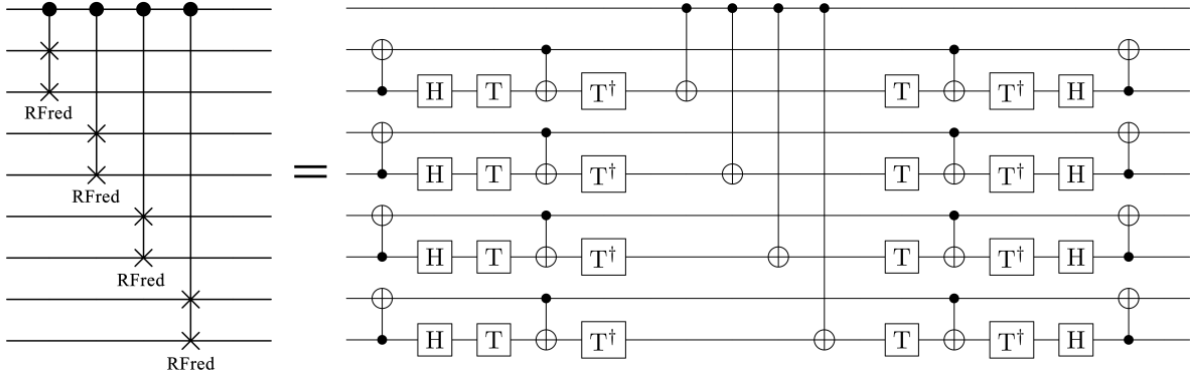

**Fig. 7** Automatic parallelization of T gates in constructing the cyclic operator when using the relative-phase Fredkin gate shown in Fig. 1(b).

**Table 1** Comparison of circuit costs for the QSM algorithm. The table presents the required resources for the QSM algorithms, including T-count (the number of T gates), T-depth (the depth of T gates), CNOT-count (the number of CNOT gates), and Qubit-count (the number of qubits). Here, $N$ represents the size of the database to be searched, and $M$ represents the size of the pattern. For the comparison, we assume that $N^{1/2}$ Grover iterations are implemented.

| Source | Ref. [4] | This work |
|---|---|---|
| T-count | $14N^{\frac{3}{2}}\log_2 N - 14N^{\frac{3}{2}}$ $+7N\log_2 N - 7N + 8N^{\frac{1}{2}}\log_2 N$ $+N^{\frac{1}{2}}(8M - 20) + 7$ $= 14N^{\frac{3}{2}}\log_2 N - O(N^{3/2})$ | $8N^{\frac{3}{2}}\log_2 N - 10N^{\frac{3}{2}}$ $+4N\log_2 N - 4N + 8N^{\frac{1}{2}}\log_2 N$ $+N^{\frac{1}{2}}(8M - 26) + 1$ $= 8N^{\frac{3}{2}}\log_2 N - O(N^{3/2})$ |
| T-depth | $5N^{\frac{1}{2}}\log_2^2 N + 9N^{\frac{1}{2}}\log_2 N$ $+N^{\frac{1}{2}}(4M + 2)$ $+\frac{5}{2}\log_2^2 N + \frac{5}{2}\log_2 N$ $= 5N^{\frac{1}{2}}\log_2^2 N + O(N^{\frac{1}{2}}\log N)$ | $4N^{\frac{1}{2}}\log_2^2 N + 8N^{\frac{1}{2}}\log_2 N$ $+N^{\frac{1}{2}}(4M - 2)$ $+2\log_2^2 N + 2\log_2 N$ $= 4N^{\frac{1}{2}}\log_2^2 N + +O(N^{\frac{1}{2}}\log N)$ |
| CNOT-count | $16N^{\frac{3}{2}}\log_2 N - 14N^{\frac{3}{2}} + 7N\log_2 N$ $-7N + 10N^{\frac{1}{2}}\log_2 N$ $+N^{\frac{1}{2}}(8M - 10) + M + 7$ $= 16N^{\frac{3}{2}}\log_2 N - O(N^{3/2})$ | $10N^{\frac{3}{2}}\log_2 N - 10N^{\frac{3}{2}} + 5N\log_2 N$ $-5N + 6N^{\frac{1}{2}}\log_2 N$ $+N^{\frac{1}{2}}(8M - 14) + M + 5$ $= 10N^{\frac{3}{2}}\log_2 N - O(N^{3/2})$ |
| Qubit-count | $\frac{3}{2}N + \log_2 N + M - 1$ | $N + \log_2 N + M$ |

# 5 Conclusion

The string-matching problem is a widespread challenge in computer science. As the scale of the problem increases, it becomes essential to enhance the efficiency of algorithms by reducing computational costs. In this paper, we have reduced the T-count of the QSM algorithm from $14N^{3/2}\log_2 N - O(N^{3/2})$ to $8N^{3/2}\log_2 N - O(N^{3/2})$. Additionally, our approach reduces other circuit costs of the QSM algorithm, including T-depth. Given that T gates dominate the cost of implementing fault-tolerant quantum circuits, our results may expedite the utilization of quantum advantages in solving string-matching problems, potentially impacting fields such as bioinformatics, text processing, and cryptography.

To achieve this cost reduction, we introduced the concept of the relative-phase Fredkin gate, drawing inspiration from the relative-phase Toffoli gate [23]. Given that relative-phase Toffoli gates are extensively used to reduce the cost of quantum circuits [23, 26−28], we anticipate that relative-phase Fredkin gates will also find broad applications in the cost reduction of other quantum circuits.

By employing the Toffoli gate implementation method presented in Ref. [29] and the Fredkin gate synthesis method presented in Ref. [24], it is possible to execute a Fredkin gate using four T gates. This method can achieve a T-count reduction comparable to our approach. However, this Fredkin gate implementation method necessitates measurement, classically controlled gates, an ancilla qubit, and additional Clifford gates. In contrast, our method, which utilizes the relative-phase Fredkin gate shown in Fig. 1(b), employs a straightforward circuit synthesis approach. This enables us to reduce the T-count without incurring additional circuit costs.

## Appendix A: cost for constructing the cyclic operator

In Appendix A, we present the exact number of Fredkin gates required to construct the cyclic operator $C$ in the QSM algorithm.

*Lemma 2.* When the size of the data to be searched is $n$, constructing $C_{2^0}$ requires $2^n - 1$ Fredkin gates.

*Proof.* It is sufficient to prove that a binary string $(a_0, a_1, a_2, \ldots, a_{2^{k+1}-1})$ can be transformed into $(a_1, a_2, ..., a_{2^{k+1}-1}, a_0)$ using $2^n - 1$ SWAP operations (Note that Fredkin gate is controlled-SWAP gate.). This statement can be proved by mathematical induction:

(1) When $n$ is 1, transforming $(a_0, a_1)$ into $(a_1, a_0)$ requires one SWAP operation.

(2) Inductive hypothesis: When $n$ is $k$, the statement is satisfied.

(3) When $n$ is $k + 1$, the string is $(a_0, a_1, a_2, ..., a_{2^{k+1}-1})$. The string is divided into two parts: $(a_0, a_1, a_2, ..., a_{2^k-1})$ and $(a_{2^k}, a_{2^k+1}, a_{2^k+2}, ..., a_{2^{k+1}-1})$. By the inductive hypothesis, each string can be transformed into $(a_1, a_2, ..., a_{2^k-1}, a_0)$ and $(a_{2^k+1}, a_{2^k+2}, ..., a_{2^{k+1}-1}, a_{2^k})$ using $2^k - 1$ SWAP operations. Then, the total string becomes $(a_1, a_2, ..., a_{2^k-1}, a_0, a_{2^k+1}, a_{2^k+2}, ..., a_{2^{k+1}-1}, a_{2^k})$. Using another SWAP operation, this string can be transformed into $(a_1, a_2, ..., a_{2^k-1}, a_{2^k}, a_{2^k+1}, a_{2^k+2}, ..., a_{2^{k+1}-1}, a_0)$. Then, the total number of SWAP operations required is $2(2^k - 1) + 1 = 2^{k+1} - 1$ ∎

*Theorem 2.* Given $n$, synthesizing operator $C_{2^k}$ requires $2^n - 2^k$ Fredkin gates.

*Proof.* It is sufficient to prove that a binary string $(a_0, a_1, a_2, ..., a_{2^{n+1}-1})$ can be transformed into $(a_{2^k}, a_{2^k+1}, a_{2^k+2}, ..., a_{2^k-1})$ using $2^n - 2^k$ SWAP operation, where additions and subtractions are modulo $2^{n+1}$ operations. This statement also can be proved by mathematical induction:

(1) When $k$ is 0, the statement is true by Lemma 2.

(2) Inductive hypothesis: When $k$ is $m$, the statement is satisfied.

(3) When $n$ is $m + 1$, the string is $(a_0, a_1, a_2, ... a_{2^{n+1}-1})$. The string is divided into two parts: $(a_0, a_2, a_4, ... a_{2^{n+1}-2})$ and $(a_1, a_3, a_5, ... a_{2^{n+1}-1})$. According to an inductive hypothesis, each string can be transformed into the correct state using $2^n - 1$ SWAP operations. Then, the total number of SWAP operations required is $2(2^n - 2^m) = 2^{n+1} - 2^{m+1}$ ∎

From Theorem 2, we prove that constructing a cyclic operator $C$ requires $N\log_2 N - N + 1$ Fredkin gates.

## Appendix B: cost for constructing the operator $R_0$

In Appendix B, we prove that the operator $R_0$ in the QSM algorithm can be an $n$-qubit operator, which is not explicitly presented in Ref. [4].

*Theorem 3.* To build $R_0$ used in the QSM algorithm, it is not necessary to use an $(n + N + M)$-qubit controlled-$Z$ gate; it is sufficient to use an $n$-qubit controlled-$Z$ gate.

*Proof.* Suppose that the second and third registers of $A^{-1}R_g A|k\rangle_n|0\rangle_N|0\rangle_M$ are $|0\rangle_N|0\rangle_M$ for an arbitrary $n$-qubit computational basis state $|k\rangle_n$. In this case, the states in the second and third registers in the QSM algorithm circuit before applying $R_0$ are always $|0\rangle_N|0\rangle_M$. This is because the operations for the QSM algorithm can be written as Eq. (B1).

$$Q^r A|0\rangle_n|0\rangle_N|0\rangle_M = (AR_0 A^{-1}R_g)^r A|0\rangle_n|0\rangle_N|0\rangle_M$$
$$= AR_0(A^{-1}R_g A)R_0(A^{-1}R_g A)R_0 \ldots (A^{-1}R_g A)R_0(A^{-1}R_g A)|0\rangle_n|0\rangle_N|0\rangle_M \quad (B1)$$

Therefore, to prove Theorem 3, it is sufficient to prove that the states of the second and third registers in $A^{-1}R_\psi A|k\rangle_n|0\rangle_N|0\rangle_M$ are $|0\rangle_N|0\rangle_M$ for an arbitrary $n$-qubit computational basis state $|k\rangle_n$. The proof is as follows: Let a function $g: \{0, 1, \ldots N - 1\} \to \{1, -1\}$ be defined by Eq. (B2).

$$g(k) = \begin{cases} 1, |P_k\rangle_M = |0\rangle_M \\ -1, |P_k\rangle_M \neq |0\rangle_M \end{cases} \quad (B2)$$

Then, state $A^{-1}S_\psi A|k\rangle_n|0\rangle_N|0\rangle_M$ evolves as Eq. (B3).

$$A^{-1}R_\psi A|k\rangle_n|0\rangle_N|0\rangle_M$$
$$= A^{-1}R_\psi \frac{1}{\sqrt{N}} \sum_{j=0}^{N-1} (-1)^{\vec{j}\cdot\vec{k}} |j\rangle_n |L_j\rangle_N |P_j\rangle_M$$
$$= A^{-1} \frac{1}{\sqrt{N}} \sum_{j=0}^{N-1} (-1)^{\vec{j}\cdot\vec{k}+g(j)} |j\rangle_n |L_j\rangle_N |P_j\rangle_M$$
$$= \frac{1}{N} \sum_{j=0}^{N-1} \sum_{l=0}^{N-1} (-1)^{\vec{j}\cdot\vec{k}+\vec{j}\cdot\vec{l}+g(j)} |j\rangle_n |0\rangle_N |0\rangle_M \quad \blacksquare \quad (B3)$$

# Acknowledgments

This work was supported by a 2024 Research Grant from the University of Seoul.

# Statements and declarations

## Conflict of Interest

The authors have no conflicts to disclose.